\documentclass[a4paper]{article}

\usepackage{17lomcon}        % Proceedings volume layout metrics
\usepackage{cite}             % Smart range citations
\usepackage{epsfig}           % Encapsulated PostScript figure inclusion

\begin{document}

%%%%%%%%%%%%%%%%%%%%%%%%%%%%%%%%%%%%%%%%%%%%%%%%%%%%%%%%%%%%%%%%%%
% The preamble of the paper
%%%%%%%%%%%%%%%%%%%%%%%%%%%%%%%%%%%%%%%%%%%%%%%%%%%%%%%%%%%%%%%%%%

\title{ASTROPHYSICAL COMPONENTS FROM PLANCK MAPS}

\author{Carlo Burigana$^{1,2,3}$ \email{burigana@iasfbo.inaf.it}, Tiziana Trombetti$^{1,2}$ \email{trombetti@iasfbo.inaf.it}, Daniela Paoletti$^{1,3}$ \email{paoletti@iasfbo.inaf.it},\\
Nazzareno Mandolesi$^{1,4,2}$ \email{mandolesi@iasfbo.inaf.it} and Paolo Natoli$^{2,5}$ \email{paolo.natoli@gmail.com} (on behalf of {\it Planck} Collaboration)}

\affiliation{$^{1}$INAF--IASF \footnote{Istituto Nazionale di Astrofisica -- Istituto di Astrofisica Spaziale e Fisica Cosmica} Bologna, Via Piero Gobetti 101, I-40129 Bologna, Italy \\
$^{2}$Dipartimento di Fisica e Scienze della Terra, Universit\`a degli Studi di Ferrara,\\
Via Giuseppe Saragat 1, I-44122 Ferrara, Italy\\
$^{3}$INFN \footnote{Istituto Nazionale di Fisica Nucleare}, Sezione di Bologna, Via Irnerio 46, I-40126, Bologna, Italy\\
$^{4}$Agenzia Spaziale Italiana, Viale Liegi 26, Roma, Italy\\
$^{5}$Agenzia Spaziale Italiana Science Data Center, \\ Via del Politecnico snc, 00133, Roma, Italy}

% You may repeat \author and \affiliation as many times as necessary!

\date{}
% Print it out!
\maketitle

%%%%%%%%%%%%%%%%%%%%%%%%%%%%%%%%%%%%%%%%%%%%%%%%%%%%%%%%%%%%%%%%%%
% The preamble of the paper
%%%%%%%%%%%%%%%%%%%%%%%%%%%%%%%%%%%%%%%%%%%%%%%%%%%%%%%%%%%%%%%%%%

\begin{abstract}
The {\it Planck} Collaboration has recently released maps of the microwave sky in both temperature and polarization.
Diffuse astrophysical components (including Galactic emissions, cosmic far infrared (IR) background, $y$-maps of the thermal Sunyaev-Zeldovich (SZ) effect)
and catalogs of many thousands of Galactic and extragalactic radio and far-IR sources, and galaxy clusters detected through the SZ effect are the main astrophysical products of the mission.
A concise overview of these results and of astrophysical studies based on {\it Planck} data is presented.
\end{abstract}

\section{Introduction}

\vskip -0.2cm

The analysis of the microwave to sub-mm sky plays a crucial role not only for the cosmological information contained in the cosmic microwave background (CMB) 
but also for the wealth of astrophysical information it provides.
The {\it Planck} Collaboration has recently released maps in temperature (resp. polarization) at nine (resp. seven) frequency bands between 30 and 857 GHz (resp. 353 GHz) 
reprocessing the data from the Low (up to 70 GHz) and High (from 100 GHz) Frequency Instruments (LFI and HFI).
To extract 
relatively bright variable sources, the diffuse Zodiacal Light Emission and solar system bodies,
a comparison between data taken at different times or maps derived from different surveys is performed,
since signal dependence on time and/or on the relative observer-source or -interplanetary dust cloud positions.
The separation of all the other astrophysical emissions and of CMB anisotropies is 
carried out on 
maps averaged over advantageous sets of surveys
applying, respectively for point-like sources and diffuse emissions, dedicated filters, possibly complemented by external catalogs, and algorithms 
exploiting frequency dependence and angular correlation, possibly in combination with templates at other frequency bands.

\section{Catalogs of sources and clusters of galaxies}

\vskip -0.2cm

\begin{figure}[t]
  \begin{minipage}{6cm}
     \centering
     \includegraphics[height =3.7cm,width=6cm]{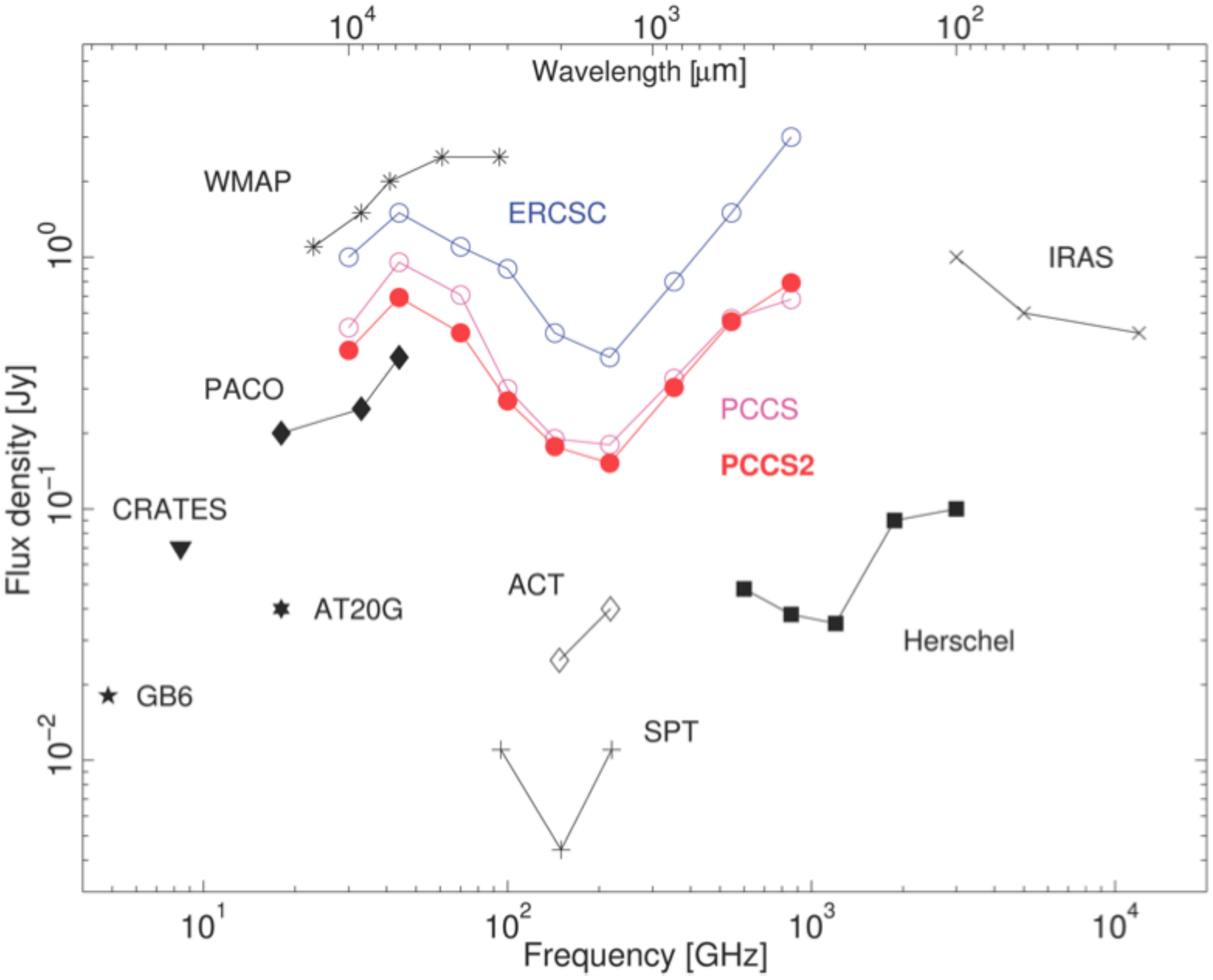}
  \end{minipage}
%\hfill
  \begin{minipage}{6cm}
     \centering
     \includegraphics[height =3.7cm,width=6cm]{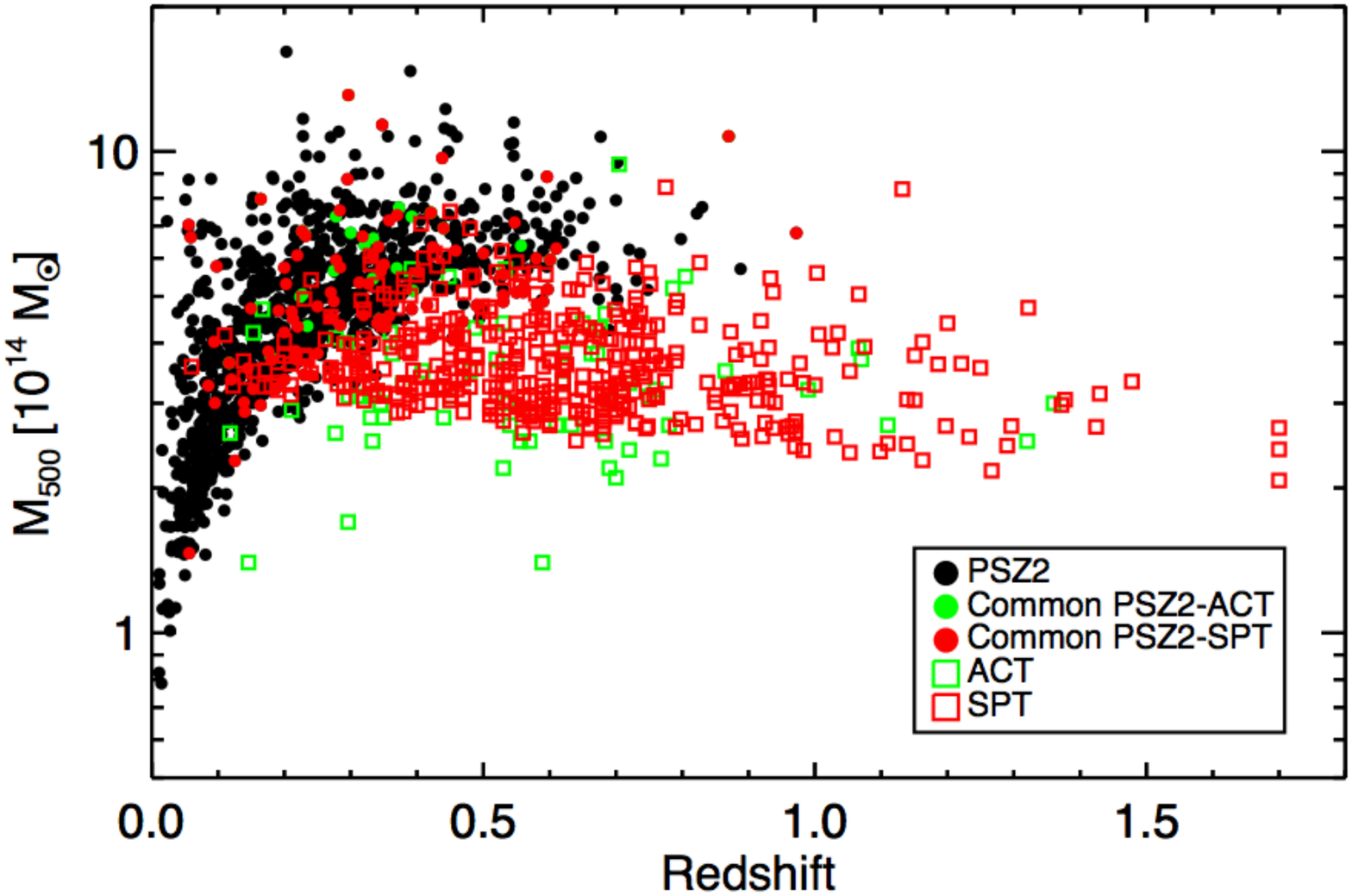}
  \end{minipage}
    \vskip -0.3cm
\caption{Left panel: sensitivity (the flux density at 90\% completeness) of the PCCS2, compared with PCCS, ERCSC, WMAP and others. 
For the LFI channels, the sensitivities refer to the full sky.
For the HFI channels, the 90\% completeness limits plotted for the PCCS were evaluated in a selected extragalactic zone;
the regions of sky to which the 90\% completeness limits apply are similar but not identical to those of the PCCS2. 
%From Ref. [\refcite{planck2015XXVI}].
From Ref. [1].
Right panel: distribution of PSZ2 clusters with associated redshift in the mass-redshift (M500 -- $z$) plane compared to SPT %(Bleem et al 2014)
and ACT %(Hasselfield et al 2013) 
catalogues. Black circles: PSZ2 clusters; red (green) filled circles: common SPT/PSZ2 (ACT/PSZ2) clusters.
Red (green) empty squares: the remaining SPT (ACT) clusters not detected by {\it Planck}. See also the text. 
%From Ref. [\refcite{planck2015XXVII}].
From Ref. [2].
}
\label{srcsz}
\vskip -0.55cm
\end{figure}

The Second {\it Planck} Catalogue of Compact Sources (PCCS2), both Galactic and extragalactic, consists of (many tens of thousands) sources detected in single-frequency maps, combined from all {\it Planck} mission data
over the entire sky, through the Mexican Hat Wavelet 2 algorithm (see Fig. \ref{srcsz}).
Four different flux-density estimates have been validated by simulations (internal validation) and comparison with other astrophysical data (external validation). 
Compact sources detected at lower (resp. higher) frequencies are assigned to the PCCS2 (resp. to the PCCS2 or PCCS2E sub-catalogues, 
depending on their sky positions). The PCCS2 covers most of the sky and allows to extract subsamples at higher reliabilities than the target 80\% integral reliability of the catalogue.
The PCCS2E contains sources whose detection reliability is affected by diffuse emission.
Both catalogs include polarized flux densities, or upper limits, and orientation angles at frequencies 30 GHz $\le \nu \le$ 353 GHz for many hundreds sources.
The spectral index proves the source nature: the low (resp. high) frequency channels are dominated by synchrotron (resp. dusty galaxies) sources. 
A significant steepening at $\nu \sim$ 50 GHz is found in blazar spectra and the bright tail of extragalactic source counts at high frequency (at least for $\nu \le 217$ GHz) is dominated 
by synchrotron emitters, not by dusty galaxies.

The Second {\it Planck} catalog of SZ galaxy clusters (PSZ2), with 1653 detections, is the largest SZ-selected sample and the deepest all-sky catalogue of galaxy clusters.
They have been analyzed with observations in other wavebands (radio to IR, optical and X-rays) and 1203 of them have counterparts in datasets external to {\it Planck}.
A population of low-redshift X-ray under-luminous clusters has been revealed by SZ selection. They appear in optical and SZ surveys with consistent mass properties,
but  are almost absent in ROSAT X-ray selected samples.
Fig. \ref{srcsz} compares PSZ2 with deeper SZ catalogs obtained with South Pole and Atacama Cosmology Telescopes (SPT and ACT) on selected areas.
Microwave and X-ray cluster data have been jointly used to independently constrain cosmological parameters. 
In particular, the evolution of galaxy cluster abundance with mass and redshift set constraints on the matter density fluctuation normalization,
the mean matter density, the dark energy density and equation of state, and on extensions of the minimal cosmological model,
including e.g. massive neutrinos and modified gravity.

Integrated information on cluster physics and cosmological evolution comes from the Comptonization parameter, $y$, fluctuations over all-sky {\it Planck} maps\cite{planck2015XXII} other than 
observing it as SZ effect towards specific clusters. The analysis of sub-millimeter/far-IR background fluctuations, carried out mainly in selected {\it Planck} 
areas, proves high-$z$ galaxy evolution and star formation history\cite{planck2013XXX}.
These themes are also probed by large scale CMB polarization, sensitive to cosmic reionization history\cite{trombettiburigana,planck2015I}
related to structure and star formation/evolution and their production of ionizing photons in the intergalactic medium.

\section{Galactic diffuse components}

\vskip -0.2cm

\begin{figure}[t]
  \begin{minipage}{5.cm}
     \centering
     \hskip -0.3cm
     \includegraphics[height =4cm,width=5cm]{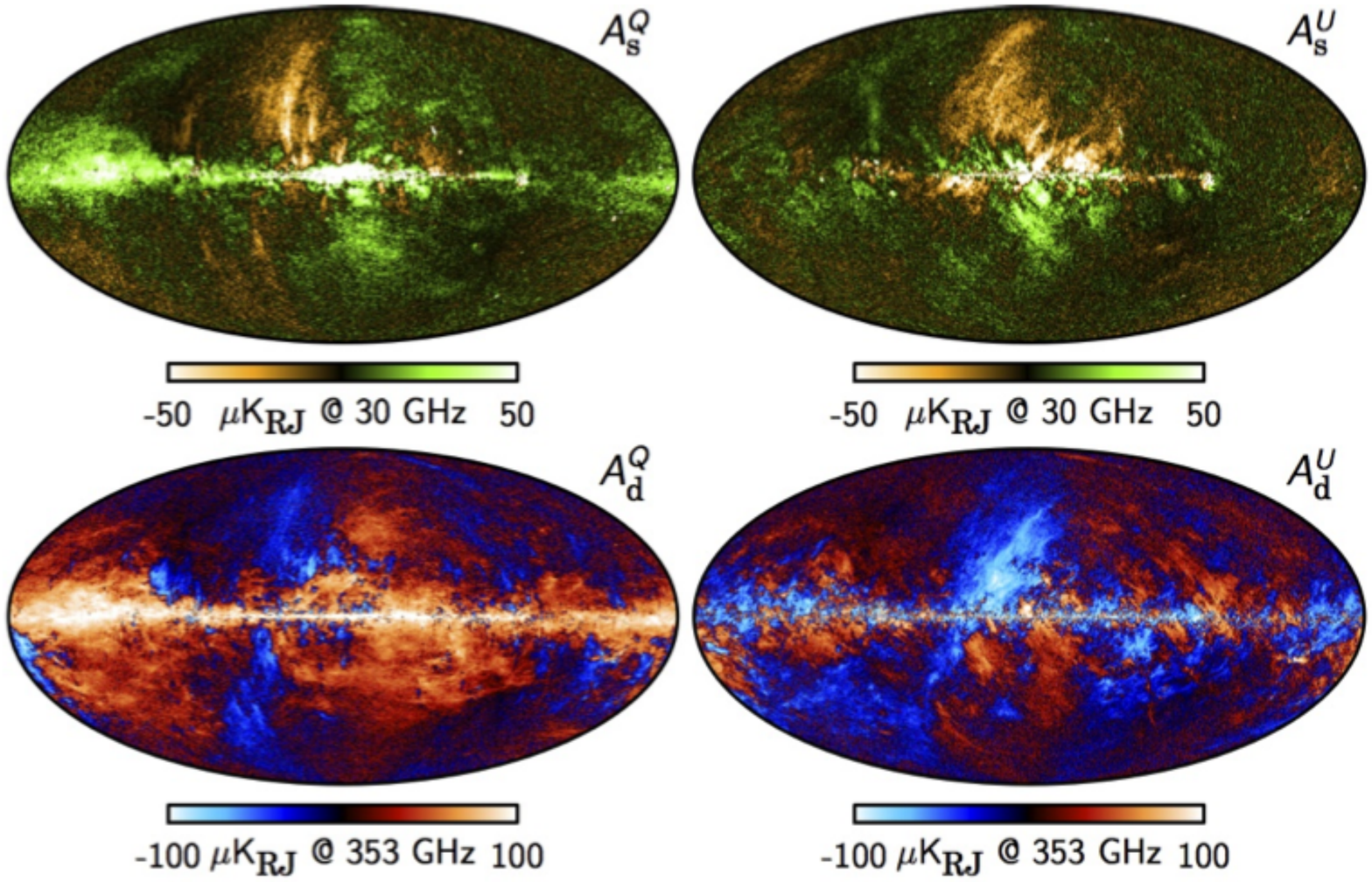}
  \end{minipage}
%\hfill
  \begin{minipage}{7.5cm}
       \hskip -0.7cm
     \centering
     \includegraphics[height =4cm,width=7.2cm]{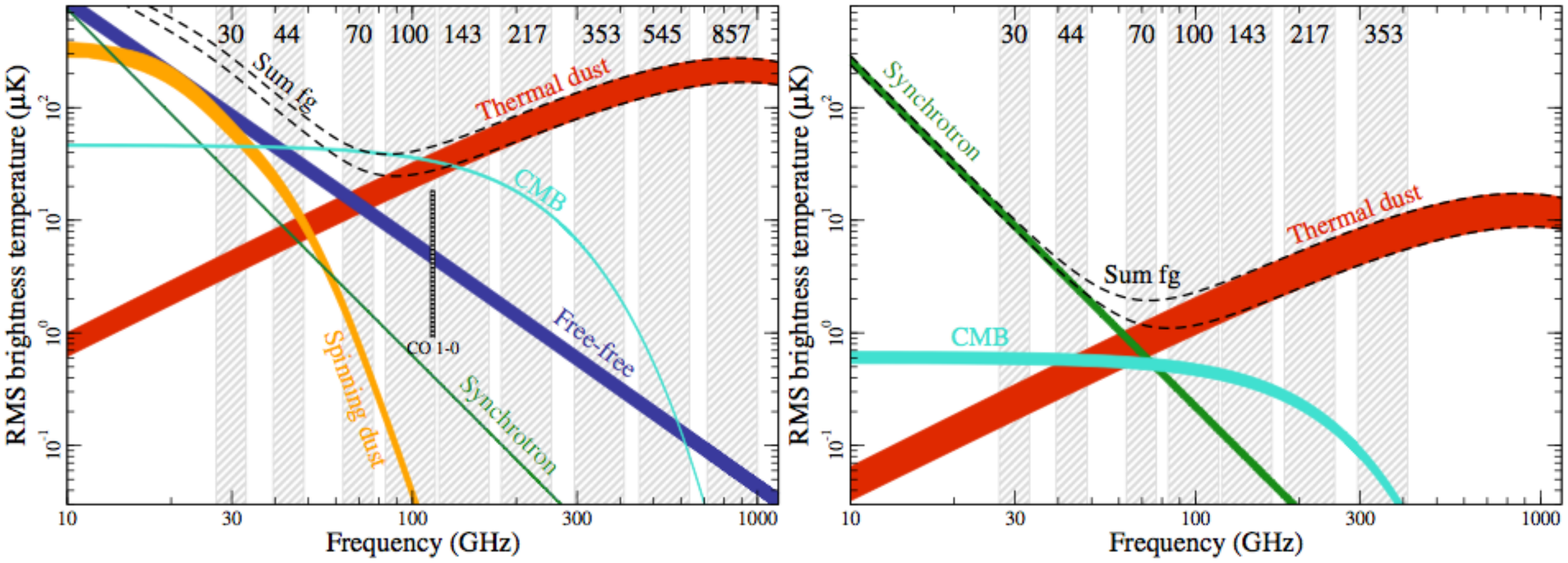}
  \end{minipage}
  \vskip -0.3cm
\caption{Maps:  Stokes Q and U parameters for the synchrotron ($s$; at 30 GHz) and thermal dust ($d$; at 353 GHz) polarization. 
Panels: brightness temperature rms as a function of frequency and component for temperature (left),
where each component is smoothed to an angular resolution of $1^{\circ}$ FWHM and the lower and upper edges of each line are defined by masks covering respectively
 81\% and 93\% of the sky, and polarization (right), where the corresponding smoothing scale is 40$'$ and the sky fractions are 73\% and 93\%. 
%From Ref. [\refcite{planck2015I}].
From Ref. [3].
}
\label{mapsrms}
\vskip -0.55cm
\end{figure}

Astrophysical diffuse components and CMB anisotropies have been separated from {\it Planck} multifrequency maps by means of four algorithms. SEVEM and COMMANDER work in real space, NILC in 
real and harmonic (needlet) space and SMICA in harmonic or needlet domain. 
The CMB maps extracted with these methods are in remarkable agreement, implying that the overall foreground emission
is analogously well set. COMMANDER, based on a-priori knowledge of foreground components characterized by parameters to be reconstructed with Bayesian methods in real space independently for each resolution element, has been found particularly suitable for characterizing astrophysical emissions varying across the sky. 
For temperature analysis, {\it Planck} data have been combined with WMAP\cite{wmap} and 
408 MHz\cite{haslam} maps
to jointly extract CMB, synchrotron, free-free, spinning dust, CO, line emission in the 94 and 100 GHz channels, and thermal dust emission. 
Fig. \ref{mapsrms} shows the Stokes Q and U parameter maps for the synchrotron and thermal dust emissions, the only diffuse astrophysical components faithfully extracted in polarization. 
Another important emission clearly appears close to the Galactic center in {\it Planck} low frequency maps:
the so-called ``haze'' component, whose full comprehension -- related to astrophysics and/or dark matter phenomena -- is still controversial.
From the various source maps it is possible to quantify the contribution of each component to sky fluctuations, displayed in Fig. \ref{mapsrms}. 
A remarkable result of {\it Planck} is the assessment of the microwave sky complexity, calling for even more frequency channels for future CMB projects aimed
at precise characterization of all components and of tiny CMB polarization features, such as primordial B-modes.
Note also the change of paradigm with respect to WMAP results in the relative weights of the various components. In temperature, synchrotron emission fluctuations are less than
those of both free-free and spinning dust emissions around 30 GHz. The angular power spectrum (APS) of synchrotron and dust polarized emissions has been properly characterized 
on various wide sky areas \cite{planck2015X} and the synchrotron frequency behavior has been studied as function of the Galactic latitude \cite{planck2015XXV}.
Polarized emission from Galactic dust, the main foreground above 70 GHz, is important (and sometimes underestimated
in previous analyses) essentially everywhere in the sky. {\it Planck} maps provided new insights into interstellar dust physics
and a precise determination of the contamination level in the CMB APS \cite{planckPIPXXX} for polarization experiments,
allowing for a proper analysis of BICEP2 and {\it Keck} array observations\cite{B2KP}.

\vskip -0.2cm

{\small

\section*{Acknowledgments}

\vskip -0.2cm
We acknowledge the use of the NASA Legacy Archive for Microwave Background Data Analysis 
and of the ESA {\it Planck} Legacy Archive.
The ASI/INAF Agreement 2014-024-R.0 for the {\it Planck} LFI Activity of Phase E2 is acknowledged.
Some results of this paper have been derived using HEALPix \cite{Gor05} package.
}

{\small

%%%%%%%%%%%%%%%%%%%%%%%%%%%%%%%%%%%%%%%%%%%%%%%%%%%%%%%%%%%%%%%%%%
% References
%%%%%%%%%%%%%%%%%%%%%%%%%%%%%%%%%%%%%%%%%%%%%%%%%%%%%%%%%%%%%%%%%%

}

\end{document}